\documentclass[11pt]{article}

\usepackage[preprint]{acl}

\usepackage{times}
\usepackage{latexsym}
\usepackage{xcolor}
\usepackage{ulem}
\usepackage{changes}

\usepackage[T1]{fontenc}

\usepackage[utf8]{inputenc}

\usepackage{microtype}

\usepackage{inconsolata}

\usepackage{graphicx}

\usepackage{booktabs}   
\usepackage{multirow}   
\usepackage{colortbl}
\usepackage{amsmath}
\usepackage{amssymb}
\usepackage{enumitem}
\usepackage{svg}
\usepackage{float}
\usepackage{pifont}     
\newcommand{\cmark}{\textcolor{green!60!black}{\ding{51}}} 
\newcommand{\xmark}{\textcolor{red!70!black}{\ding{55}}}   

\title{Adaptive Candidate Retrieval with Dynamic Knowledge Graph Construction for Cold-Start Recommendation}

\author{
    Wooseong Yang$^{1}$, Weizhi Zhang$^{1}$, Yuqing Liu$^{1}$, Yuwei Han$^{1}$, \\
    \textbf{Yu Wang}$^{1}$\textbf{,} \textbf{Junhyun Lee}$^{2}$\thanks{Corresponding authors.}\textbf{,} \textbf{Philip~S.~Yu}$^{1*}$ \\
    $^{1}$University of Illinois Chicago \quad
    $^{2}$Korea University \\
    }

\begin{document}
\maketitle

\begin{abstract}

The cold-start problem remains a critical challenge in real-world recommender systems, as new items with limited interaction data or insufficient information are frequently introduced.
Despite recent advances leveraging external knowledge such as knowledge graphs (KGs) and large language models (LLMs), recommender systems still face challenges in practical environments. Static KGs are expensive to construct and quickly become outdated, while LLM-based methods depend on pre-filtered candidate lists due to limited context windows. 
To address these limitations, we propose ColdRAG, a retrieval-augmented framework that dynamically constructs a knowledge graph from raw metadata, extracts entities and relations to construct an updatable structure, and introduces LLM-guided multi-hop reasoning at inference time to retrieve and rank candidates without relying on pre-filtered lists.
Experiments across multiple benchmarks show that ColdRAG consistently outperforms strong seven baselines. Our implementation is available at \url{https://github.com/WooseongYang/ColdRAG}.

\end{abstract}

\section{Introduction} 

In real-world recommender systems, cold-start items are routinely introduced with few or no interaction records and incomplete metadata. This lack of information prevents models from accurately estimating user preferences, resulting in poor recommendation quality, reduced user satisfaction, and ultimately revenue loss \cite{huang2023aligning,zhang2025cold}.
To tackle this challenge, recent works have explored two main directions: (i) KG–based methods that construct structured representations of the item catalog \cite{wang2019kgat,guo2020survey}, and (ii) LLM-based methods that leverage LLMs as training-free recommenders, typically prompting on user histories with a small set of candidate items \cite{sanner2023large,hou2024large}.

However, both directions face critical limitations in practical deployment \cite{lin2025can}. 
Static KGs are expensive to construct and maintain, and they quickly become outdated as items, attributes, and relations change \cite{wang2019multi}. 
Updating these graphs to reflect catalog changes typically requires substantial offline engineering and cannot keep up with real-time changes. 
LLM-based approaches, meanwhile, are usually formulated as re-rankers over a pre-filtered candidate set rather than as end-to-end retrieval systems \cite{hou2024large}. 
Because LLMs operate under bounded context windows and token budgets, only a curated subset of items and a truncated user history can be included in the prompt, necessitating a separate task-specific retrieval pipeline. While recent work introduces retrieval-aware prompting, its retrieval remains shallow, limited to keyword matching or single-hop similarity search, which still limits performance and reduces adaptability in dynamic cold-start settings \cite{liang2025taxonomy,kieu2024keyword}.


To address the above limitations, we introduce ColdRAG, a retrieval-augmented framework built around two key modules.
The first module, Dynamic Knowledge Graph Construction, automatically builds and incrementally updates a domain graph from catalog fields (e.g., titles, descriptions, attributes, reviews), allowing the structure to evolve naturally as the catalog changes.
The second module, Adaptive Candidate Retrieval over Knowledge Graph, removes the need for pre-filtered candidate lists by treating candidate generation as LLM-guided, goal-directed traversal over the graph, assembling a compact, high-utility candidate set together with evidence paths that justify each recommendation. Empirically, ColdRAG consistently surpasses strong training-based and training-free baselines across diverse product domains with large performance gains.


Our contributions are threefold:
\begin{itemize}
\item We introduce a dynamic KG construction that automatically builds and incrementally updates the graph as items and relations evolve.
\item We eliminate the unrealistic assumption that a curated candidate list is already in the LLM’s context window by integrating candidate retrieval with LLM-guided multi-hop reasoning.
\item We demonstrate strong cold-start performance on multiple benchmarks and provide extensive analyses on component effectiveness, stability, and robustness. 
\end{itemize}

\section{Related Works}
\begin{figure*}[t]
\vskip 0.15in
\begin{center}
\centerline{\includegraphics[width=\textwidth]{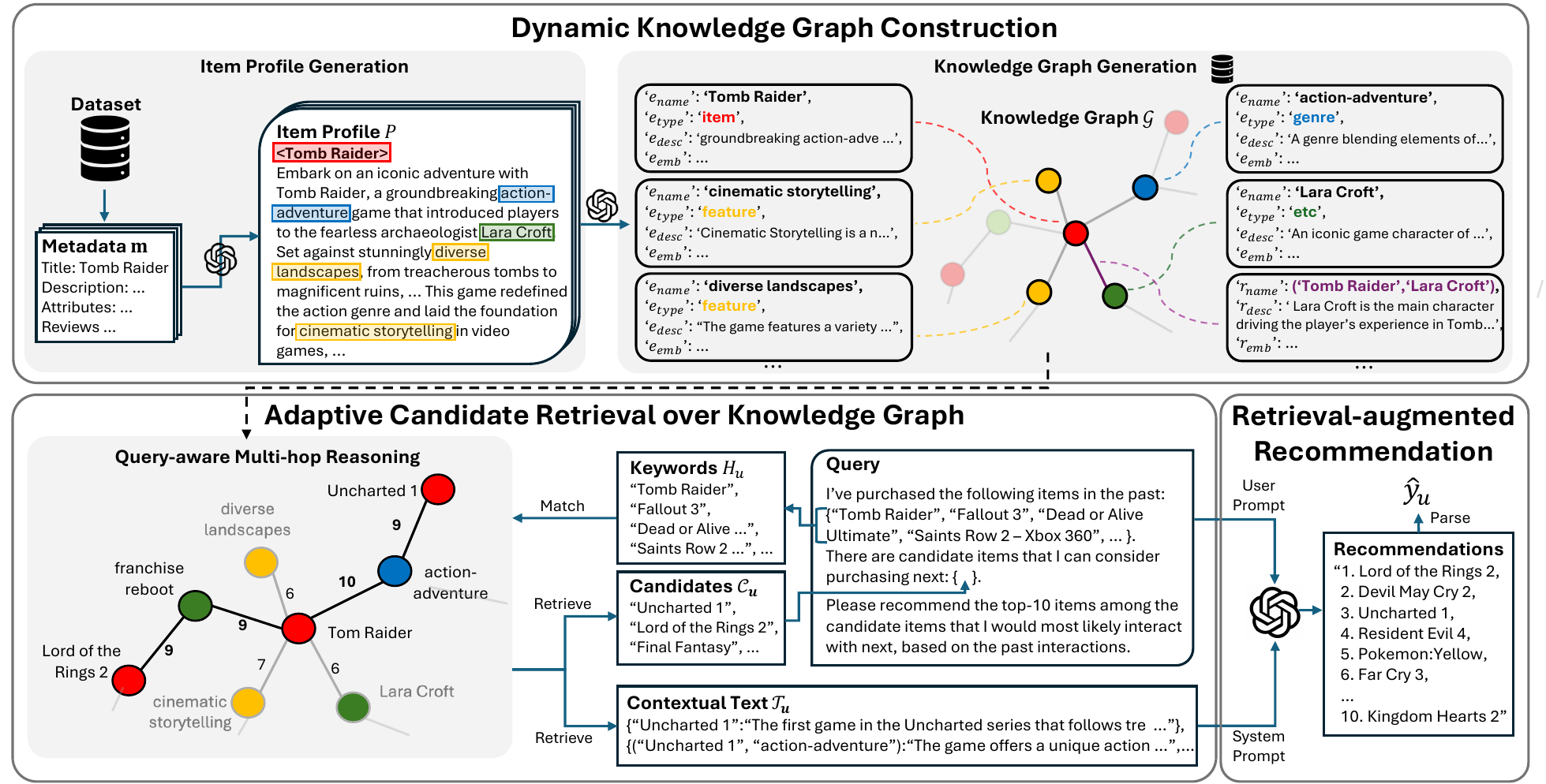}}


\caption{Overview of the proposed \textbf{ColdRAG} framework. 
Given item metadata, an LLM generates item profiles, 
from which structured entities and relations form a knowledge graph dynamically. 
During inference, ColdRAG performs query-aware multi-hop reasoning over KG 
to adaptively retrieve candidate items and context, then composes prompts to generate recommendations.}

\label{fig:framework}
\end{center}
\vskip -0.1in
\end{figure*}

\paragraph{Cold-Start Recommendation}

The item cold-start problem arises when new items lack interaction history, motivating content-based and hybrid recommenders that rely on metadata or auxiliary signals to compensate for missing collaborative information. 
Representative training-based methods such as CLCRec \cite{wei2021contrastive} strengthen cold-start representations through contrastive alignment, while TDRO \cite{lin2024temporally} improves robustness by accounting for temporal distribution shifts.
Although effective, these approaches require training task-specific modules, limiting their flexibility when new items continually appear.

\paragraph{LLM-based Recommendation}
To address these limitations, recent work has explored training-free LLM-based recommendation as a natural fit for cold-start scenarios. 
Methods such as LLMRank \cite{hou2024large}, TaxRec \cite{liang2025taxonomy}, and KALM4Rec \cite{kieu2024keyword} infer user preferences through prompting or lightweight retrieval, allowing them to handle new items without finetuning. 
However, due to the limited context window, these models rely on pre-filtered candidate lists or shallow keyword matching, which restricts semantic coverage and increases the risk of hallucination and unstable outputs.
ColdRAG overcomes these issues by dynamically constructing a KG and performing multi-hop, evidence-grounded retrieval, providing structured semantic grounding that enables more reliable and adaptable zero-shot item cold-start recommendation.

\section{Proposed Method}
\label{sec:method}

We present \textbf{ColdRAG}, a retrieval-augmented generation framework for cold-start recommendation. 
ColdRAG equips an LLM with a dynamically constructed KG that enables semantic reasoning and adaptively builds candidate items for context-aware and controllable recommendation. The overall framework is illustrated in Figure~\ref{fig:framework}.
For the problem setting, we follow the sequential recommendation task, where each user \(u\) has an interaction history \(H_u = [i_1, i_2, \dots, i_{n-1}]\) and the goal is to recommend the next item \(i_n\).

\subsection{Dynamic Knowledge Graph Construction}
\subsubsection{Item Profile Generation}

Item metadata is a key source for constructing the knowledge graph, but it is often sparse, noisy, or inconsistently structured, making it difficult to extract meaningful semantics directly.
To address this, ColdRAG leverages a LLM to denoise and enrich this information by using the model’s pretrained knowledge to fill informational gaps, producing concise, knowledge-grounded item profiles that capture each item’s essential semantics.
For each item \( i \), we define its metadata as \( \mathbf{m}_i = (\text{title}, \text{description}, \text{attributes}, \text{review}) \) and obtain the profile via the inline mapping \( P_i = \textit{LLM}(\textit{prompt}(\mathbf{m}_i)) \).
This process curates raw, unstructured metadata into fluent summaries that highlight key concepts such as \textit{genre}, \textit{features}, or \textit{notable entities} (e.g., “action-adventure,” “Lara Croft”), forming a clean and standardized foundation for knowledge graph construction and downstream reasoning. The prompt used for this step is provided in Appendix~\ref{sec:appendix_promptA}.

\subsubsection{Knowledge Graph Generation}

With enriched item profiles in place, ColdRAG organizes this information into a structured semantic graph by prompting the LLM to extract entities and relations, which creates a foundation for reasoning and retrieval.
Given an item profile \( P_i \), the LLM extracts entities, relations, and textual statements describing them, constructing a knowledge graph \( \mathcal{G} = (\mathcal{E}, \mathcal{R}) \), where \( \mathcal{E} \) denotes entities and \( \mathcal{R} \) the relations linking them. 

To be specific, for each item, the LLM extracts entities and relations, each paired with a natural-language description and an embedding for semantic retrieval.
An entity \( e \in \mathcal{E} \) is represented as \( (e_{\text{name}}, e_{\text{type}}, e_{\text{desc}}, e_{\text{emb}}) \), where \(e_{\text{name}}\) is the entity title, \(e_{\text{type}}\) identifies its category (e.g., item, genre, feature), \(e_{\text{desc}}\) provides its textual explanation, and \(e_{\text{emb}}\) is its vector embedding. 
A relation \( r \in \mathcal{R} \) is represented as \( (r_{\text{name}}, r_{\text{desc}}, r_{\text{emb}}) \), where \(r_{\text{name}} =(e_{\text{src}}, e_{\text{tgt}})\) denotes the source-target entity pair, \( r_{\text{desc}} \) describes their connection, and \( r_{\text{emb}} \) is its embedding used for scoring. 
For instance, as shown in Figure~\ref{fig:framework}, the entity Tomb Raider is of type 'item' and the description “groundbreaking action-adventure game …”.
A corresponding relation connects Tomb Raider to Lara Croft with the description “Lara Croft is the main character driving the player’s experience in Tomb Raider.”.
This transformation organizes free-form text into an entity-centric graph structure, enabling fine-grained multi-hop reasoning over attribute-level connections. The prompt used is shown in Appendix~\ref{sec:appendix_promptB}.

The constructed graph is stored in a hybrid knowledge base: the graph topology (nodes and edges) and their textual descriptions are stored as structured files, while embeddings are indexed in a vector database (e.g., \textit{FAISS}\footnote{\url{https://github.com/facebookresearch/faiss}}) for efficient similarity search. 
Each textual description is encoded using the same pretrained embedding model, ensuring consistent semantic representations across all entities and relations; additional details are provided in Appendix~\ref{sec:appendix_additional}.
This combination of structural traversal and semantic retrieval provides ColdRAG with evidence-grounded access to item knowledge during recommendation.

\subsection{Adaptive Candidate Retrieval over Knowledge Graph}
Once the knowledge graph \( \mathcal{G} \) has been constructed, ColdRAG uses it to adaptively identify candidate items aligned with a user's current interests. 
Given a user query that includes both task instructions and the interaction history \( H_u \), the system begins by locating the parts of the graph most relevant to the user. 
The titles of items in \( H_u \) are used as keyword anchors and embedded with the same pretrained embedding model from graph construction.
These embeddings are then matched against stored entity embeddings using cosine similarity to locate the most semantically similar nodes. 
The matched entities initialize the frontier \( \mathcal{F}_0 \), representing the user's current semantic context in the graph.
ColdRAG then performs iterative query-aware multi-hop reasoning guided by the LLM. 
At each step \( t \), all outgoing edges from the current frontier \( \mathcal{F}_t \) are scored by the LLM according to their relevance to the user history, where \( s_r = \textit{LLM}(r_{\text{desc}}, H_u) \) and \( s_r \in [0,10] \) measures the semantic alignment between the relation description \( r_{\text{desc}} \) and the user's interests. 
Edges with \( s_r \ge \lambda \) are retained, and their target nodes form the next frontier:
\[
\mathcal{F}_{t+1} = \{ e' \mid (e, e', r_{\text{desc}})\!\in\!\mathcal{R},\ s_r\!\ge\!\lambda \}.
\]
When a target node corresponds to an item, it is added to a temporary candidate pool \( \widetilde{\mathcal{C}}_u \) along with its associated descriptions \( \widetilde{\mathcal{T}}_u \).  
Traversal continues until \(|\widetilde{\mathcal{C}}_u|\) reaches the predefined maximum pool size \(\theta_{\text{pool}} \). 
Finally, the LLM aggregates edge scores to rank the retrieved items and selects the top \( \theta_{\text{top}} \) as the final candidate set \( \mathcal{C}_u \), with their textual evidence forming the final contextual input \( \mathcal{T}_u \). The example prompt is shown in Appendix~\ref{sec:appendix_promptC}.

\subsection{Retrieval-augmented Recommendation}

In the final stage, ColdRAG generates recommendations using the candidate item set \( \mathcal{C}_u \) and contextual text block \( \mathcal{T}_u \). 
The contextual text, composed of natural-language descriptions of relevant entities and relations from the knowledge graph, serves as the \textit{system prompt} that provides semantic grounding. 
The candidate set is integrated into the user query to form the \textit{user prompt} \( Q_u \), which expresses user preferences and specifies the desired top-\(k\) recommendations within the retrieved candidates.  

The LLM then generates ranked outputs conditioned on both prompts:
\[
\hat{\mathcal{Y}}_u = \textit{ParseTopK}(\textit{LLM}(\mathcal{T}_u, Q_u, k)).
\]
Here, \( \textit{LLM}(\mathcal{T}_u, Q_u, k) \) denotes generation with top-\(k\) instruction (e.g., ``Recommend the top-\(k\) items among the given candidate list, based on the user's history and retrieved context''), and \( \textit{ParseTopK} \) extracts top-k ranked item titles from the output. ColdRAG’s ability to accommodate new items is further discussed in Appendix~\ref{sec:appendix_cold}.

\begin{table}[ht]
\centering
\caption{
Summary of dataset and constructed knowledge graph statistics.
}
\resizebox{\linewidth}{!}{
\begin{tabular}{lrrrrrr}
\toprule
\textbf{Dataset} & \textbf{\#Interactions} & \textbf{\#Items} & \textbf{\#Users} & \textbf{\#Nodes} & \textbf{\#Edges} \\
\midrule
Games  & 45,106  & 2,027  & 2,096  & 15,048  & 29,023 \\
Toys   & 332,055 & 12,342 & 20,390 & 58,096  & 132,229 \\
Office & 233,738 & 6,107  & 15,302 & 42,769  & 75,053 \\
\bottomrule
\end{tabular}
}
\label{tab:statistics}
\end{table}

\begin{table*}[ht]
\centering
\footnotesize
\setlength\tabcolsep{3pt}
\caption{
Comparison of Recall@10 and NDCG@10 (\%) across three datasets. 
Our proposed ColdRAG is highlighted in gray. 
Best results are in bold and second-best baseline is underlined. 
All results are \textbf{averaged over 5 runs} and values are shown as \textbf{percentages}.
}
\begin{tabular}{lllcccccccc}
\toprule[1.5pt]
& \multirow{2.5}{*}{\textbf{Model}} 
& \multirow{2.5}{*}{\textbf{LLM}} 
& \multicolumn{2}{c}{\textbf{Games}} 
& \multicolumn{2}{c}{\textbf{Toys}} 
& \multicolumn{2}{c}{\textbf{Office}} \\
\cmidrule(lr){4-5} \cmidrule(lr){6-7} \cmidrule(lr){8-9}
& & & \textbf{Recall@10} & \textbf{NDCG@10}
  & \textbf{Recall@10} & \textbf{NDCG@10}
  & \textbf{Recall@10} & \textbf{NDCG@10} \\
\midrule[1pt]

\multirow{3}{*}{\shortstack{\textbf{training}\\\textbf{-based}}}
& UniSRec & – & 1.14 & 0.48 & 1.48 & 0.71 & 1.41 & 0.66 \\
& CLCRec  & – & 5.71 & 2.98 & 2.75 & 1.36 & 3.36 & 1.98 \\
& TDRO    & – & 6.61 & 4.21 & 2.64 & 1.31 & 3.79 & 2.14 \\
\midrule[1pt]

\multirow{12}{*}{\raisebox{-45pt}{\shortstack{\textbf{training}\\\textbf{-free}}}}
& LLM              & GPT  & 3.48 & 1.52 & 1.32 & 0.60 & 3.16 & 1.34 \\
&                   & Qwen & 9.24 & 3.89 & 0.56 & 0.33 & 3.30 & 1.63 \\ \cmidrule(lr){2-9}
& LLMRank (S)      & GPT  & 4.62 & 1.94 & 1.36 & 0.58 & 3.80 & 1.63 \\
&                   & Qwen & 10.92 & 5.14 & 1.01 & 0.65 & 3.67 & 1.87 \\ \cmidrule(lr){2-9}
& LLMRank (R)      & GPT  & 8.86 & 4.25 & 1.13 & 0.72 & \underline{4.20} & 2.23 \\
&                   & Qwen & \underline{10.98} & \underline{5.39} & 1.20 & 0.71 & 2.27 & 1.42 \\ \cmidrule(lr){2-9}
& LLMRank (I)      & GPT  & 6.20 & 3.78 & 1.20 & 0.70 & 4.02 & 1.90 \\
&                   & Qwen & 9.08 & 4.88 & 1.24 & 0.87 & 4.07 & \underline{2.39} \\ \cmidrule(lr){2-9}
& TaxRec           & GPT  & 3.75 & 1.78 & 0.60 & 0.34 & 2.20 & 0.96 \\
&                   & Qwen & 7.61 & 4.63 & 0.80 & 0.59 & 3.81 & 2.12 \\ \cmidrule(lr){2-9}
& KALM4Rec         & GPT  & 8.50 & 4.14 & \underline{4.26} & \underline{2.13} & 3.43 & 1.71 \\
&                   & Qwen & 7.27 & 2.62 & 3.29 & 1.90 & 2.07 & 0.58 \\ \cmidrule(lr){2-9}
& \cellcolor{gray!15}\textbf{ColdRAG} 
                   & \cellcolor{gray!15}GPT
                   & \cellcolor{gray!15}12.38 & \cellcolor{gray!15}4.37
                   & \cellcolor{gray!15}\textbf{5.40} & \cellcolor{gray!15}\textbf{2.29}
                   & \cellcolor{gray!15}8.60 & \cellcolor{gray!15}3.26 \\
& \cellcolor{gray!15}                   
                   & \cellcolor{gray!15}Qwen
                   & \cellcolor{gray!15}\textbf{19.57} & \cellcolor{gray!15}\textbf{6.50}
                   & \cellcolor{gray!15}4.10 & \cellcolor{gray!15}1.98
                   & \cellcolor{gray!15}\textbf{9.40} & \cellcolor{gray!15}\textbf{3.92} \\
\midrule[0.5pt]
\multicolumn{3}{c}{\textbf{Improvement}} 
& 78.22\% & 20.54\% & 26.76\%  & 7.42\% & 123.81\% & 63.85\% \\
\bottomrule[1.5pt]
\end{tabular}
\label{tab:main_results}
\end{table*}

\section{Experiments}



We conduct comprehensive experiments to evaluate the effectiveness of ColdRAG in item cold-start recommendation. 
Our analysis is organized around the following research questions:

\begin{itemize}[leftmargin=*, itemsep=1pt, topsep=0pt, parsep=0pt, partopsep=0pt]
    \item \textbf{RQ1:} Does ColdRAG effectively address the item cold-start recommendation problem?
    \item \textbf{RQ2:} How effective is the Dynamic Knowledge Graph Construction?
    \item \textbf{RQ3:} How does Adaptive Candidate Retrieval enhance recommendation performance?
    \item \textbf{RQ4:} Does ColdRAG exhibit stable and consistent generation across runs?
    \item \textbf{RQ5:} Does ColdRAG reduce hallucination and avoid out-of-domain recommendations?
\end{itemize}

\subsection{Experimental Setup}

\subsubsection{Datasets}


We evaluate ColdRAG on three domains from the Amazon Review dataset~\cite{ni2019justifying}: \textit{Games}, \textit{Toys}, and \textit{Office}, which represent diverse product types and interaction patterns. 
We apply core filtering with a threshold of 15 for \textit{Games} and 10 for \textit{Toys} and \textit{Office}, retaining users and items that meet the minimum interaction count. 
To simulate item cold-start scenarios, the least frequent 10\% of items in each dataset are designated as \textit{cold items}. 
Following the sequential recommendation setting, each user’s interactions are treated as a sequence, where the last item is held out for testing under the leave-one-out protocol~\cite{sun2019bert4rec,hou2022towards}. 
From each domain, we sample 500 user sequences that end with a cold item; the preceding \( n{-}1 \) items serve as input and the final item as the test target. 
For training-based baselines, these 500 cold-item sequences are used for testing, and all remaining sequences form the training set, ensuring consistent evaluation between training-based and training-free settings. 
Dataset and knowledge graph statistics are summarized in Table~\ref{tab:statistics}.

\subsubsection{Baselines}



We compare ColdRAG with representative baselines spanning both \textit{training-based} and \textit{training-free} paradigms.
Among training-based models, \textbf{UniSRec}~\cite{hou2022towards} fine-tunes a universal sequence encoder with contrastive objectives, \textbf{CLCRec}~\cite{wei2021contrastive} trains a contrastive framework to preserve collaborative signals for cold items, and \textbf{TDRO}~\cite{lin2024temporally} applies distributionally robust optimization to handle temporal shifts.
For training-free methods, we include a plain \textbf{LLM} that ranks randomly sampled candidates without retrieval grounding; \textbf{LLMRank}~\cite{hou2024large}, which provides sequential (S), recency (R), and in-context (I) prompting variants for zero-shot re-ranking; \textbf{TaxRec}~\cite{liang2025taxonomy}, which injects taxonomy cues to guide LLM reasoning; and \textbf{KALM4Rec}~\cite{kieu2024keyword}, which uses keyword-level retrieval to support cold-start recommendation.
Among the \textit{training-based} baselines, UniSRec fine-tunes a pretrained model, whereas CLCRec and TDRO are trained from scratch. 
The remainings are \textit{training-free} methods that operate entirely without parameter updates, relying on LLM inference for reasoning and retrieval. 
Together, they provide a broad comparison across representation learning, fine-tuning, and retrieval-augmented LLM paradigms.



\subsubsection{Evaluation Metrics}

We evaluate recommendation performance using two standard metrics widely adopted in cold-start tasks: Recall@\(k\) and NDCG@\(k\), following prior work~\cite{hou2022towards,wei2021contrastive,liang2025taxonomy}. 
All results are reported at \( k = 10 \).  


\subsubsection{Implementation Details}

ColdRAG and all training-free baselines are implemented using two LLMs: \textit{gpt-4o-mini}\footnote{\url{https://platform.openai.com/docs/models/gpt-4o-mini}} and \textit{Qwen2.5-32b-instruct}\footnote{\url{https://huggingface.co/Qwen/Qwen2.5-32B-Instruct}} (“GPT” and “Qwen” in Table~\ref{tab:main_results}). 
Using two distinct LLM backbones verifies that ColdRAG’s effectiveness generalizes beyond a single architecture. 
We set the edge scoring threshold to \( \lambda = 7 \), the candidate pool size to \( \theta_{\text{pool}} = 300 \), and the final candidate set size to \( \theta_{\text{top}} = 100 \), consistent with Section~\ref{sec:method}. 
All experiments are repeated five times, and average results are reported for stability and reproducibility. 
Additional implementation details appear in Appendix~\ref{sec:appendix_additional}.

\subsection{Results}

\subsubsection{Overall Performance (RQ1)}

Table~\ref{tab:main_results} compares ColdRAG with all baselines on the \textit{Games}, \textit{Toys}, and \textit{Office} datasets. ColdRAG consistently achieves the best results across all metrics and domains, outperforming both \textit{training-based} and \textit{training-free} baselines. It improves Recall@10 by +78.6\%, +26.76\%, and +123.81\% over the strongest baselines on \textit{Games}, \textit{Toys}, and \textit{Office}, respectively. The larger gains in Recall over NDCG indicate that ColdRAG is effective in both identifying and ranking relevant items, but its main strength lies in retrieving correct candidates into the top set.
Among \textit{training-based} models, TDRO shows the best performance but still falls short of ColdRAG, showing that even robustly trained models struggle to generalize under sparsity. ColdRAG’s retrieval-augmented design instead captures fine-grained semantic relations through LLM-guided multi-hop reasoning, yielding better cold-start adaptability.
Within the \textit{training-free} methods, LLM and LLMRank perform moderately but rely on fixed candidate lists, restricting contextual exploration. KALM4Rec enriches prompts via keyword retrieval yet remains less strong than ColdRAG, whose KG-based retrieval enables deeper reasoning over semantically linked concepts beyond shallow keyword matching.
Overall, ColdRAG’s consistent gains underscores the benefit of integrating KG-based retrieval with LLM reasoning for robust item cold-start recommendation.

\begin{figure}[t]
\vskip 0.1in
\centering
\includegraphics[width=\linewidth]{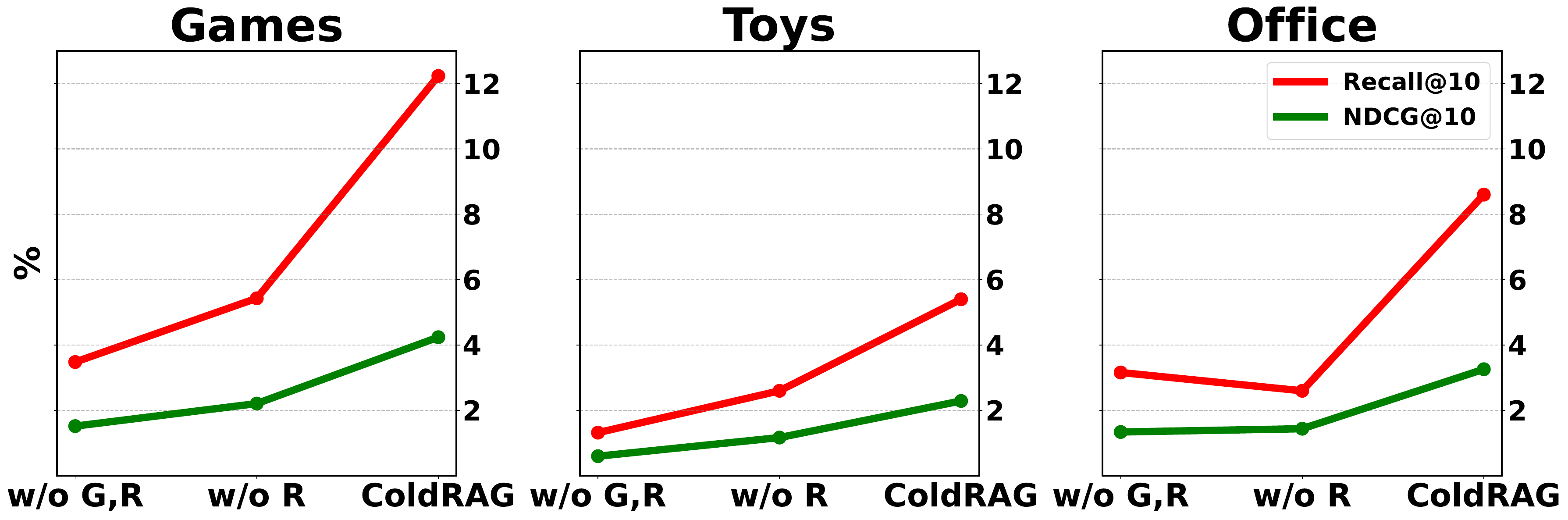}
\caption{Performance comparison of ColdRAG variants across three domains using GPT, showing that both core modules (G and R) add performance gains.}
\label{fig:ablation}
\vskip -0.1in
\end{figure}

\subsubsection{Ablation Study (RQ2 \& RQ3)}

To examine the impact of ColdRAG’s core components, we compare three settings: \textbf{w/o G,R}, a plain LLM without dynamic knowledge graph construction (\textbf{G}) or adaptive candidate retrieval (\textbf{R}); \textbf{w/o R}, a variant that includes G but replaces R with embedding-similarity top-k matching; and the full ColdRAG model combining both modules.
As shown in Figure~\ref{fig:ablation}, performance improves steadily from \textbf{w/o G,R} to \textbf{ColdRAG} across most domains. These results indicate that dynamic knowledge graph construction (\textbf{G}) provides structured semantic grounding, while adaptive candidate retrieval (\textbf{R}) introduces goal-directed exploration over related entities, together yielding consistent gains as each module is added. In the \textit{Office} domain, \textbf{w/o R} performs slightly worse than \textbf{w/o G,R}, indicating that unfiltered knowledge can introduce noise when metadata is sparse. However, the full model restores performance by selectively refining relevant information through reasoning. 
Overall, the two modules are complementary: knowledge grounding provides semantic depth, and retrieval ensures relevance.
Their combination drives ColdRAG’s superior cold-start recommendation performance.

\subsubsection{Analysis on Stability and Hallucination (RQ4 \& RQ5)}

\begin{figure}[t]
\vskip 0.15in
\centering
\includegraphics[width=\linewidth]{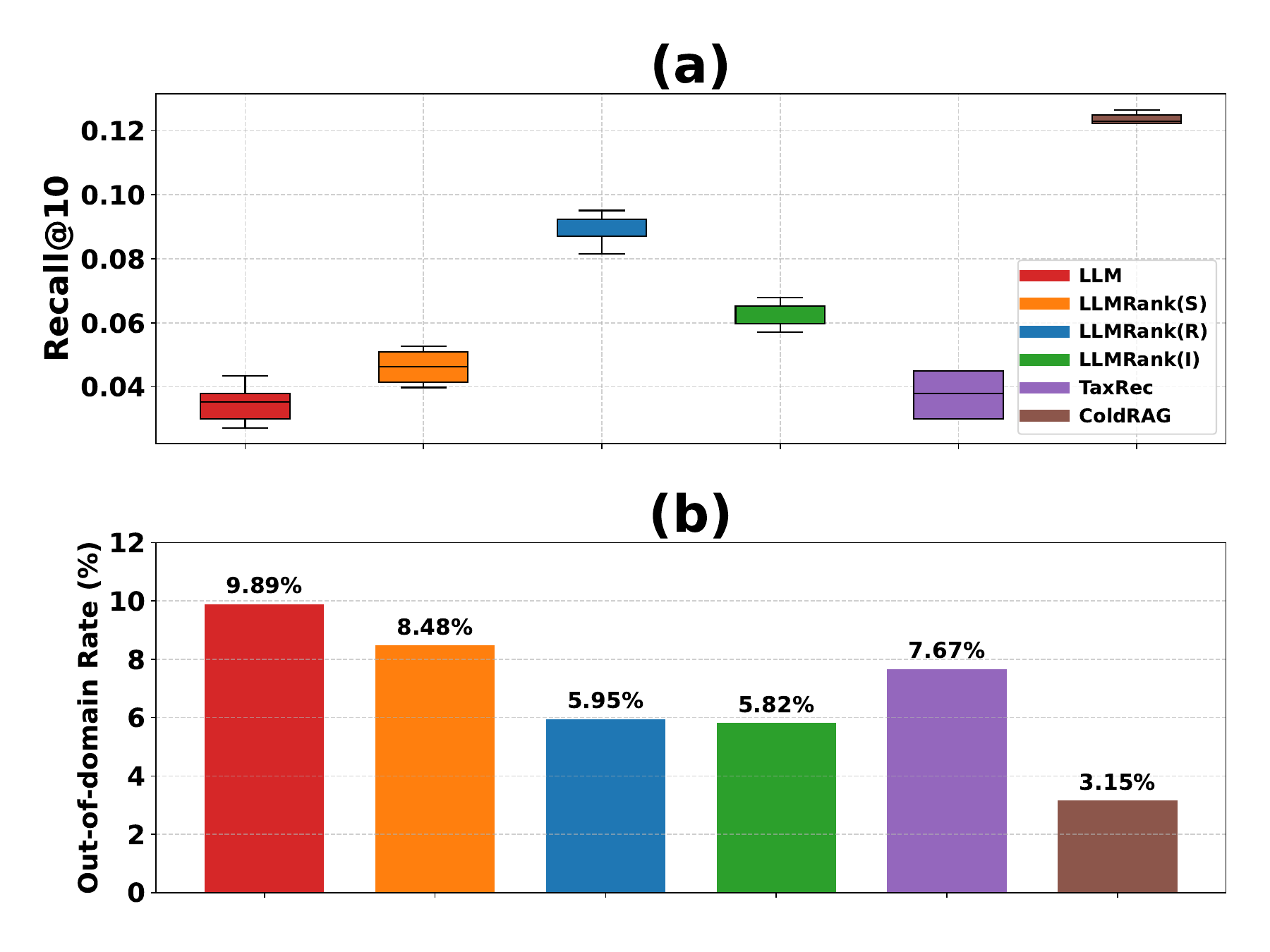}
\caption{(a) Recall@10 box plots over five runs for training-free baselines. (b) Out-of-domain generation rates, both evaluated on the \textit{Games} dataset with GPT.}
\label{fig:ood_stability}
\vskip -0.1in
\end{figure}

LLM-based recommenders often suffer from generation inconsistency and hallucination, producing unstable or out-of-domain outputs. We evaluate ColdRAG on these aspects using five independent runs on the \textit{Games} dataset. As shown in Figure~\ref{fig:ood_stability}, ColdRAG achieves high average performance and low variance in Recall@10, demonstrating stable, reproducible generation compared to other training-free baselines. This robustness stems from structured retrieval and reasoning, providing consistent semantic grounding rather than relying on prompt randomness.
We also measure hallucination rates, defined as the proportion of generated items not present in the dataset. While other LLM-based models, including LLM and LLMRank variants, exhibit 5–10\% out-of-domain outputs even with predefined candidate lists, ColdRAG reduces this rate to 3.15\%. This improvement shows that knowledge-grounded retrieval helps the model construct and reason over a semantic graph, enabling it to retrieve contextually valid items and constrain generation within the domain. 
In summary, beyond achieving superior recommendation performance, ColdRAG also exhibits robust stability and minimal hallucination, which are essential for practical and trustworthy recommender systems.

\section{Conclusion}

We presented ColdRAG, a retrieval-augmented generation framework for item cold-start recommendation. ColdRAG dynamically builds a knowledge graph from sparse metadata and performs LLM-guided multi-hop reasoning to adaptively retrieve candidate items aligned with user preferences—without relying on pre-built candidate lists. This design enables accurate and stable recommendations, making ColdRAG a practical and industry-ready solution for real-world cold-start scenarios.


\section*{Limitations}

While ColdRAG demonstrates strong performance, it faces several practical constraints. First, its reliance on repeated LLM queries during knowledge graph construction and multi-hop reasoning introduces notable computational cost and latency, posing challenges for large-scale or real-time deployment. Moreover, although ColdRAG can operate with both open- and closed-source LLMs, reliance on closed-sourced models such as the GPT series can make reproduction costly and less consistent across environments.
Another limitation lies in ColdRAG’s limited adaptability. Several key hyperparameters, such as edge scoring thresholds and candidate pool sizes, are manually set and static across domains. This rigidity may constrain performance under varying data distributions or interaction sparsity. A more adaptive, agentic framework could dynamically adjust these parameters and query strategies, improving both efficiency and generalization in diverse real-world settings.

\appendix


\section{Cold-start Adaptability}
\label{sec:appendix_cold}

\begin{figure}[H]
\vskip 0.1in
\centering
\includegraphics[width=\linewidth]{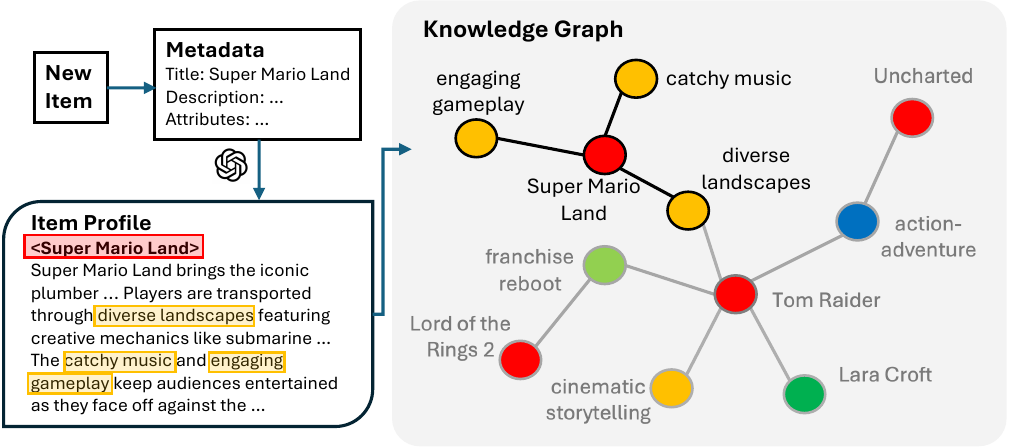}
\caption{Illustration of ColdRAG’s adaptability to item cold-start scenario.}
\label{fig:colditem}
\vskip -0.1in
\end{figure}

ColdRAG is inherently suitable for item cold-start scenarios, as illustrated in Figure~\ref{fig:colditem}. 
When a new item appears with only metadata, the framework immediately generates its item profile, extracts entities and relations, and integrates them into the existing knowledge graph through the same pipeline used for prior items. 
This enables the new item to connect with semantically related concepts (e.g., genres, features, or characters) and become part of the graph’s reasoning and retrieval processes without requiring historical interactions or retraining. 
Such seamless integration allows ColdRAG to remain robust and responsive in dynamic environments where new content is frequently introduced.

\section{Comparison of Recommender System Categories}
\label{sec:appendix_comparison}

Traditional recommender systems, including collaborative filtering (CF), content-based (CB), and hybrid CF+CB models, often struggle when new items lack interaction data. CF depends on dense user–item histories and weakens under sparsity, while CB relies on incomplete or noisy metadata. Hybrid methods partially alleviate these issues but still require user interactions to anchor predictions. LLM-based recommenders introduce a zero-shot alternative by leveraging pretrained knowledge, yet prompt-only designs remain vulnerable to hallucination, instability, and limited coverage. RAG improves grounding through external evidence but often overlooks relational structure or retrieves overly broad neighborhoods. ColdRAG addresses these limitations by performing structured multi-hop reasoning over a dynamically constructed knowledge graph, producing evidence-grounded, zero-shot recommendations that are robust to cold-start scenarios. As shown in Table~\ref{tab:comparison}, ColdRAG is the only framework that jointly supports cold-start handling, external grounding, zero-shot usability, and multi-hop reasoning.

\begin{table}[H]
\centering
\caption{Comparison of recommender paradigms across four key capabilities: handling cold-start, grounding in external evidence, zero-shot usability, and multi-hop reasoning over structured knowledge. (* denotes static metadata rather than retrieved evidence.}
\resizebox{1.0\columnwidth}{!}{
\begin{tabular}{lcccc}
\toprule
& \textbf{Cold-start} & \textbf{Grounding} & \textbf{Zero-shot} & \textbf{Multi-hop} \\
\midrule
CF & \xmark & \xmark & \xmark & \xmark \\
CB           & \cmark & \cmark\textsuperscript{*} & \xmark & \xmark \\
CF + CB        & \cmark & \cmark & \xmark & \xmark \\
Prompt-only LLM         & \cmark & \xmark & \cmark & \xmark \\
RAG-based LLM         & \cmark & \cmark  & \cmark & \xmark \\
\textbf{ColdRAG (ours)} & \cmark & \cmark & \cmark & \cmark \\
\bottomrule
\end{tabular}
}
\label{tab:comparison}
\end{table}

\section{Structure of Knowledge Graph}
\label{sec:appendix_structure}
\begin{figure}[H]
\vskip 0.1in
\centering
\includegraphics[width=\linewidth]{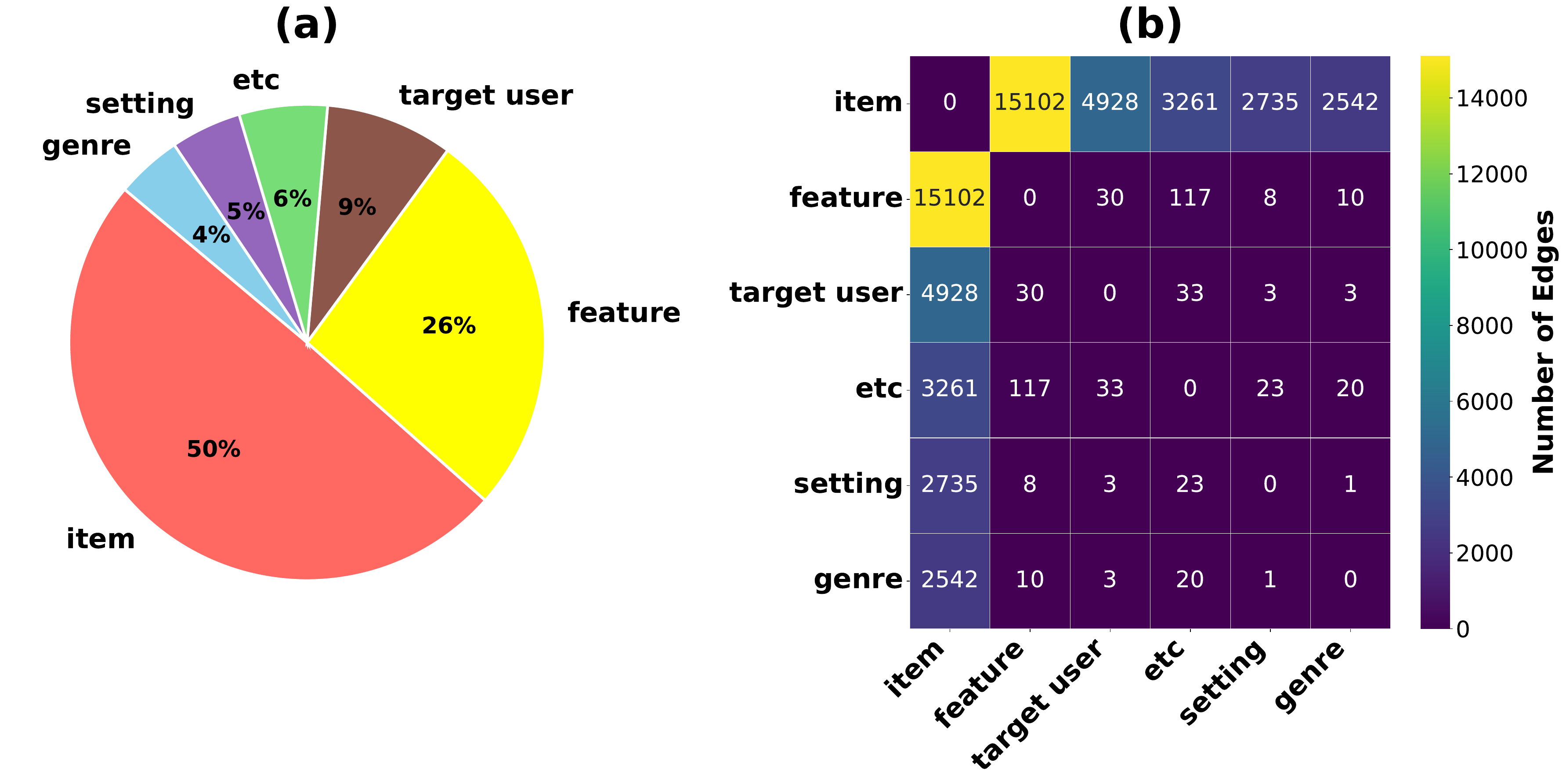}
\caption{(a) Pie chart of the distribution of entity (node) types in the KG. (b) Heatmap of the number of relations (edges) between the entity types.}
\label{fig:kg_analysis}
\vskip -0.1in
\end{figure}

We analyze the structure of the knowledge graph (KG) generated using the \textit{Games} dataset with \textit{gpt-4o-mini}, focusing on the distribution of entity types and their relations, as shown in Figure~\ref{fig:kg_analysis}. The KG is primarily composed of \textit{item} nodes (50\%) and \textit{feature} nodes (26\%), while other entities such as \textit{target user} (9\%), \textit{etc} (6\%), \textit{setting} (5\%), and \textit{genre} (4\%) provide complementary semantic context. This distribution shows that the KG is centered around items, with non-item entities describing and explaining their properties. The heatmap in Figure~\ref{fig:kg_analysis} reveals dense connections between items and features (15,102 edges) and between items and target users (4,928 edges), indicating that the graph effectively captures item attributes and user-related semantics. Additional links to genre and setting nodes further enrich contextual diversity, enabling nuanced multi-hop reasoning. Overall, the KG exhibits a dense yet interpretable structure that supports ColdRAG’s retrieval and reasoning processes.

\section{Prompt Templates}

We present the prompt templates used in the four core modules of ColdRAG.  
Each prompt is designed to guide the LLM through a distinct stage of the pipeline, ensuring consistent and interpretable behavior.

\subsection{Item Profile Generation}
\label{sec:appendix_promptA}
This prompt directs the LLM to create a concise, fluent item profile using the title, metadata, and reviews, enriching sparse information with its pretrained knowledge when necessary.
\begin{figure}[H]
\vskip 0.15in
\centering
\includegraphics[width=\linewidth]{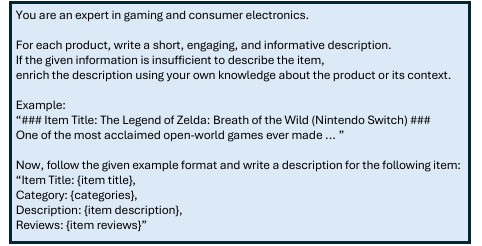}
\caption{Example prompt for Item Profile Generation.}
\label{fig:promptA}
\vskip -0.1in
\end{figure}

\subsection{Dynamic Knowledge Graph Construction}
\label{sec:appendix_promptB}
This prompt instructs the LLM to extract entities and relations from the generated item profile, producing a structured and interpretable knowledge graph centered around the item.
\begin{figure}[H]
\vskip 0.15in
\centering
\includegraphics[width=\linewidth]{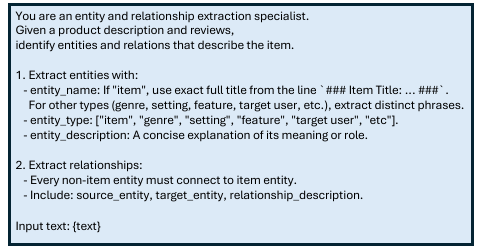}
\caption{Example prompt for Entity and Relation Extraction.}
\label{fig:promptB}
\vskip -0.1in
\end{figure}

\subsection{Adaptive Candidate Retrieval over KG}
\label{sec:appendix_promptC}
This prompt enables the LLM to evaluate graph edges using the user’s interaction history and iteratively expand the reasoning frontier to identify semantically relevant candidate items.
\begin{figure}[H]
\vskip 0.15in
\centering
\includegraphics[width=\linewidth]{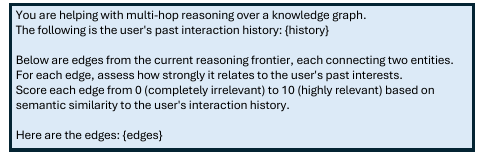}
\caption{Example prompt for Adaptive Candidate Retrieval.}
\label{fig:promptC}
\vskip -0.1in
\end{figure}

\subsection{Retrieval-augmented Generation}
\label{sec:appendix_promptD}
This prompt guides the LLM to rank the retrieved candidate items and produce the final top-\(k\) recommendations in a dataset-consistent format.
\begin{figure}[H]
\vskip 0.15in
\centering
\includegraphics[width=\linewidth]{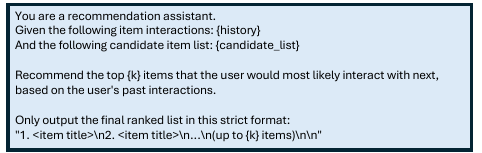}
\caption{Example prompt for Retrieval-augmented Generation.}
\label{fig:promptD}
\vskip -0.1in
\end{figure}

\section{Hyperparameter Analysis}
\begin{figure}[H]
\vskip 0.1in
\centering
\includegraphics[width=\linewidth]{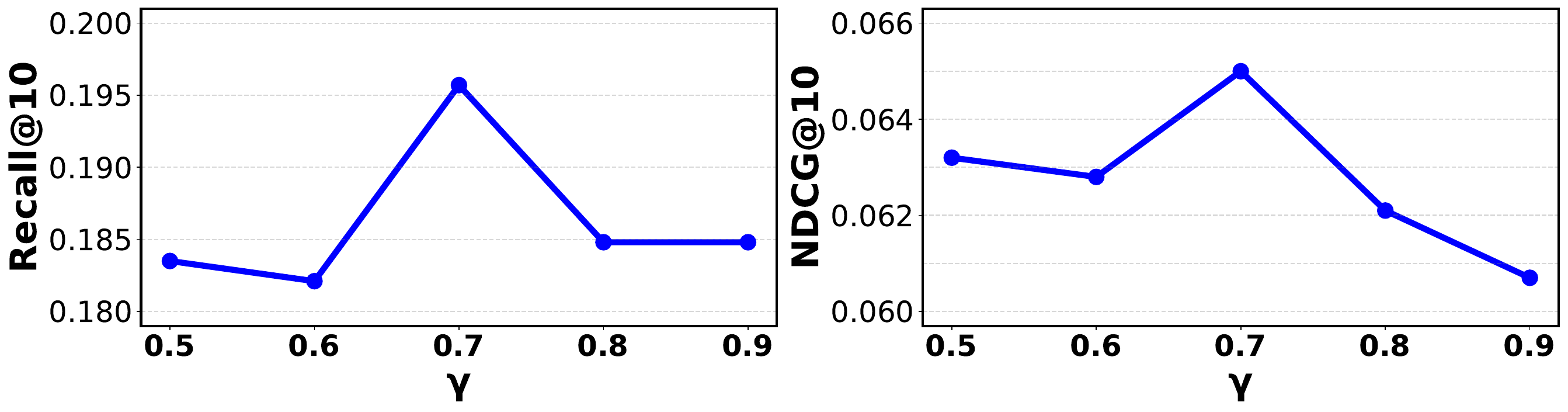}
\caption{Effect of the edge scoring threshold \( \lambda \) on ColdRAG’s performance.}
\label{fig:hyperparam}
\vskip -0.1in
\end{figure}

We analyze the impact of the edge scoring threshold \( \lambda \), which controls how strictly ColdRAG filters edges during multi-hop reasoning. 
As shown in Figure~\ref{fig:hyperparam}, ColdRAG achieves the best performance when \( \lambda = 0.7 \). 
A smaller threshold allows irrelevant edges to remain, while an excessively large threshold overly constrains traversal and misses useful nodes. 
This result indicates that a moderate threshold effectively balances relevance and diversity in the retrieved candidates, yielding the most robust overall performance.

\section{Additional Implementation Details}
\label{sec:appendix_additional}
For the GPT setting, we use \textit{gpt-4o-mini} accessed through the Azure OpenAI API. 
Entity and relation embeddings are encoded using OpenAI’s \textit{text-embedding-3-small} model, with all embeddings indexed in \textit{FAISS} for approximate nearest-neighbor retrieval. 
For the Qwen setting, we employ \textit{qwen2.5-32b-instruct} served via the \textit{vLLM}\footnote{\url{https://github.com/vllm-project/vllm}} backend, paired with the \textit{bge-m3}\footnote{\url{https://huggingface.co/BAAI/bge-m3}} embedding model for semantic representation. 
Both configurations follow identical hyperparameters and retrieval settings to ensure a fair comparison across LLM backbones. 
These results confirm that ColdRAG’s performance is consistent across different LLM architectures, demonstrating its architecture-agnostic robustness.




\end{document}